\documentstyle[aps]{revtex}

\begin{document}

\draft

\title{\bf Characterization of the Local Density of States Fluctuations \\
near the Integer Quantum Hall Transition in a Quantum Dot Array }

\author{ Giancarlo Jug\cite{GJ}$^a$ and Klaus Ziegler$^{a,b}$ }

\address{ $^a$Max-Planck-Institut f\"ur Physik Komplexer Systeme, Au\ss
enstelle Stuttgart, Postfach 800665, D-70569 Stuttgart (Germany) }

\address{ $^b$Institut f\"ur Physik, Universit\"at Augsburg, D-86135 Augsburg
(Germany)}

\date{\today}

\maketitle

\begin{abstract} We present a calculation for the second moment of the local
density of states in a model of a two-dimensional quantum dot array near the
quantum Hall transition. The quantum dot array model is a realistic adaptation
of the lattice model for the quantum Hall transition in the two-dimensional
electron gas in an external magnetic field proposed by Ludwig, Fisher,
Shankar and Grinstein. We make use of a Dirac fermion representation for the
Green functions in the presence of fluctuations for the quantum dot energy
levels. A saddle-point approximation yields non-perturbative results for the
first and second moments of the local density of states, showing interesting
fluctuation behaviour near the quantum Hall transition.  To our knowledge we
discuss here one of the first analytic characterizations of chaotic behaviour
for a two-dimensional mesoscopic structure.  The connection with possible
experimental investigations of the local density of states in the quantum dot
array structures (by means of NMR Knight-shift or single-electron-tunneling
techniques) and our work is also established.  \end{abstract}

\pacs{PACS numbers: 71.55.Jv, 73.20.Dx, 73.40.Hm, 05.30.-d}

\section{ Introduction }

Recently, there has been a considerable surge of interest in the
electronic properties of nanometric-scale metal and semiconductor structures
\cite{alw,cks}. Static as well as transport properties of these systems can be
obtained from the knowledge of the statistics of the energy level distribution
for quantum-mechanical structures in which weak disorder (or quantum chaos)
plays a fundamental role. The presence of an external magnetic field enhances
the influence of the quantum fluctuations on the properties of these mesoscopic
systems. The remarkable advances in nanostructure fabrication procedures have
also lead recently to a good deal of experimental and theoretical studies of
the properties of Quantum Dots, Quantum Wires and, moreover, of Quantum Dot
Arrays (QDA) \cite{weiss} which can be obtained by a variety of techniques at
semiconductor surfaces and interfaces. All these mesoscopic structures exhibit
a wealth of new interesting quantum phenomena in an external magnetic field
and at low temperatures. In particular, QDA (as ``artificial crystals'') offer
the possibility of investigating situations not accessible in natural crystals
for the magnetic fields that can be produced in normal laboratory conditions.

On the other hand, a most celebrated quantum phenomenon, occurring at low
temperatures in the presence of a magnetic field and assisted by the presence
of weak disorder in a two-dimensional (2D) semiconductor heterostructure,
is the Integer Quantum Hall Effect (IQHE) \cite{kli}. The plateaux in the
Hall conductivity are believed to be the result of current-carrying edge
states and localized bulk states \cite{halperin}.  The transitions between
plateaux are due to localization-delocalization transitions induced by the
combined effects of the magnetic field and disorder close to zero absolute
temperature \cite{reviews}. The behaviour near a single IQHE transition
(QHT) can be described in a number of ways, theoretically, in order to
reproduce the jump in the conductivity and other singularities near the
QHT. A model for the IQHE that is particularly suited to the present work
has been proposed and studied by Ludwig et al. \cite{lud}. In this model,
a square lattice hopping model is defined, with a fixed half-flux-quantum
$\frac{1}{2}\Phi_0$ per plaquette defining the lattice spacing $a$, with both
nearest-neighbour and next-nearest-neighbour hopping processes considered
in the presence of a staggered chemical potential and suitable forms of the
one-particle disordered potential. In the continuum limit, $a\rightarrow
0$ formally, and near the QHT this model yields a valid description of the
behaviour near a single IQHE step \cite{lud,zie1}.

It would seem natural, therefore, to extend the model of Ludwig et al. to
the description of the QHT in a QDA system. This has indeed been done
\cite{zie2}, by taking into account the fluctuations of the quantum levels
pertinent to a single quantum dot. The existence of a supporting lattice
in the QDA also makes the model of Ludwig et al., as extended and studied
by Ziegler \cite{zie2}, physically more appealing. In this model a QHT is
indeed found and described in the proximity of the transition point. It is
found that the otherwise sharp QHT step becomes rounded by the presence of
level-fluctuations in the single-dots.

In this paper we present an analytical characterization of the fluctuations
of the local density of states (DOS) for a QDA near the QHT, using the model
and calculations presented in \cite{zie2}. Our calculation is interesting
in its own right for a good number of reasons. First and foremost, the
averaged DOS and the moments of its distribution due to level statistics
can be studied by means of Knight shift NMR measurements, now becoming
accessible for semiconductor heterostructures \cite{ks}. Another candidate
technique for measuring local DOS-related properties is the single-electron
tunneling (SET) \cite{schmidt}, which provides an imaging of the local DOS in
(bulk) semiconductors. A more detailed discussion will be presented in the
Conclusions, Section IV. Also, from the more theoretical point of view, it is
of interest to characterise the fluctuations near the QHT of an observable
quantity like the DOS, considering that in the absence of fluctuations
in the levels of a dot the DOS itself vanishes at the centre of the band,
$E=0$. When fluctuations are present, a semi-circle law is obtained for the
averaged local DOS $\langle \rho({\bf r}) \rangle$ which does not vanish
anymore at $E=0$. The second moment of the global DOS vanishes identically,
but not that of the local DOS

\begin{equation} M_2(m)=\langle \rho({\bf r},m)^2 \rangle - \langle \rho({\bf
r},m) \rangle^2, \end{equation}

\noindent which we calculate in this work. Here, $m$ plays the role of a
chemical potential which controls the density of fermions in the system
(for details see Section II). We find that $M_2$ diverges near the point
$m=m_c$ where $\langle \rho({\bf r},m) \rangle$ vanishes, indicating that the
level-fluctuations are particularly strong near this value of the energy. From
the technical point of view, our work represents an attempt to characterise
quantum mesoscopic fluctuations in a system of dimensionality $D > 1$ without
applying the $2+\epsilon$ expansion technique. Most approximate methods in
fact deal with $D=0$ systems in the end (e.g. single dots, or aggregates of
small metallic particles) and more rigorous and/or numerical work has been
done chiefly for one-dimensional systems. An alternative approach to random
electronic systems remains, naturally, the above-mentioned $2+\epsilon$
expansion technique \cite{wegn,prsc}, which however applies only to the
scaling regime near the mobility edge.  Therefore, our contribution to
this field is instrumental for extending theoretical research to systems
of dimensionality $D=2$, as our calculations can indeed be implemented,
with some additional computational effort, for the characterisation of the
higher moments of the local DOS, as well as of those for the conductivity.
As a further remark, we point out that the straight Quantum Hall system can be
viewed as a mesoscopic-like system itself in terms of the recently-discovered
\cite{coko} universal conductance fluctuations between Quantum Hall plateaux
near the QHT. This indicates, as confirmed by very recent numerical studies
\cite{wjl,chfi}, that the characterization of the electronic states in the
presence of level fluctuations for 2D electron-gas systems is of interest
even before the complications of a periodic QDA structure are inserted in
these devices.

The article is organised as follows. In Section II we define the details
of the model and justify its applicability to the QDA system in the close
proximity of the QHT. This is a model for quasiparticles (quasielectrons) in
a lattice of `` artificial atoms'' (the QDA); the quasiparticle levels are
subject to some statistics with strong correlations of universal nature and
characterized by a Wigner-Dyson distribution \cite{wig,mehta}. The model is
set up with a fixed flux $\frac{1}{2}\Phi_0$ per plaquette, but although this
may imply a fixed value of the magnetic field we stress that the presence
of a (staggered) chemical potential $\mu$ allows one to tune the model in
the close vicinity of a QHT. The model includes a random potential, and
we take the point of view that a random (Dirac) mass for the electrons is
adequate in capturing the physics of the level fluctuations inside a single
quantum dot. In Section III we briefly discuss the functional integral
representation for the appropriate products of Green functions, following
standard procedures \cite{efet,zie3}. We set up a representation in terms of
superfields (bosonic as well as fermionic) to take care of the averaging over
level fluctuations, and carry out a large-$N$ saddle-point approximation to
evaluate the first and second moments of the local DOS near the QHT. This is a
justified approximation when $N\approx 10^2$ represents the number of single
quantum dot levels, but the case $N=1$, representing the straight Quantum
Hall system, can in principle also be described by this approximation.
We carry out the analysis for the Gaussian fluctuations and present our
results. These are discussed in the conclusions, Section IV, where the
connection with some other recent papers on the subject is established,
as is the connection between our results and their possible experimental
verification by means of NMR Knight shift or SET measurements.

\section{ The Model }

A 2D array of quantum dots in a homogeneous perpendicular magnetic field
can be modelled by a tight-binding tunneling (or hopping) Hamiltonian
which, neglecting quasiparticle-interactions, takes the quadratic form
${\cal H}=\sum H_{r,r'}^{\alpha,\alpha '}c_r^{\alpha\dagger}c_{r'}^{\alpha
'} +{\rm h.c.}$, with $c_r^{\alpha\dagger}$ creating a quasielectron at
QDA site $r$ in single-dot level $\alpha=1,2,\dots,N$. The matrix elements
$H_{r,r'}^{\alpha,\alpha '}$ are chosen as \cite{zie2}

\begin{equation} H_{r,r'}^{\alpha,\alpha '}=H_r^{(0)\alpha,\alpha
'}\delta_{r,r'} +H_{r,r'}^{(t)}\delta_{\alpha,\alpha '}
+V_r\delta_{r,r'}\delta_{\alpha,\alpha '}.  \label{lattmodel} \end{equation}

\noindent Here we take the point of view that the single-dot
electronic states are statistically distributed with matrix-elements
$H_r^{(0)\alpha,\alpha '}$ defined by the Gaussian unitary ensemble (GUE)
statistics: $\langle H_r^{(0)\alpha,\alpha '} H_r^{(0)\beta,\beta '} \rangle=
(g/N)\delta_{\alpha,\beta}\delta_{\alpha ',\beta '}$, $g$ being the strength
of the level fluctuations depending on the nature of the interactions and/or
disorder and chaos inside a single dot. We take, at least initially, the point
of view that weak tunneling processes take place between neighbouring dots and
within the same single-dot energy levels. Strictly-speaking our formulation
already contains tunneling between different neighbouring dots' energy
levels via the random occupation of levels pertinent to each individual dot.
Here, tunneling diagonal in the level indices will be assumed as realistic for
the actual calculations and we take the tunneling rates to be non-vanishing
only between nearest-neighbouring and next-nearest-neighbouring dots. These
rates take into account the presence of a magnetic field that can be thought
of as fixed at the value $B=\Phi_0/2a^2$, in correspondence with the spacing
$a$ (typically \cite{weiss} one has $a=100$ to 500 nm) between the dots. In
the Landau gauge, the tunneling Hamiltonian matrix elements read (setting
${\bf r}=(x,y)$ and indicating with $e_x$ and $e_y$ the lattice unit vectors):

\begin{equation} H_{r,r'}^{(t)}=te^{\pi i
y/a}\delta_{r',r+e_x}+t\delta_{r',r+e_y} \pm it'e^{\pi i
y/a}\delta_{r',r+e_x{\pm}e_y} + {\rm h.c.}.  \label{tunnhamil} \end{equation}

\noindent Finally, the potential $V_r$ represents an additional (e.g. electric)
external field and can be regarded as a staggered chemical potential:
$V_r=(-1)^{x+y}\mu$, which opens a gap $2\mu$ in the spectrum of the
quasielectrons. For the ``simple'' IQHE situation, the model reduces to its
$N=1$ limit and corresponds to the model proposed for the QHT by Ludwig et
al. \cite{lud}. We assume for the QDA that the separation $a$ between the
dots is much greater than the dots size.

Now, following a standard procedure \cite{lud,zie1}, the model
(\ref{lattmodel}) is reduced in the continuum limit (formally $a\rightarrow 0$)
and for long wavelengths to an equivalent 2D Dirac Hamiltonian \cite{zie1,zie2}

\begin{equation}
H^{\alpha\beta}=(\sigma_1\nabla_1+\sigma_2\nabla_2)\delta_{\alpha,\beta}
+\sigma_3 M^{\alpha\beta}\ \ \ \ (\alpha,\beta=1,...,N).  \label{diracham}
\end{equation}

\noindent $\nabla$ is here the 2D (lattice) gradient operator, $\sigma_j$ are
Pauli matrices and $M$ is a random mass matrix with mean $m\equiv\mu-t'$:
$M_r^{\alpha\beta}=m\delta_{\alpha,\beta}+\delta M_r^{\alpha\beta}$ with
$\langle\delta M_r^{\alpha\beta}\delta M_{r'}^{\alpha'\beta'} \rangle_{\delta
M}=(g/N)\delta_{\alpha,\beta'}\delta_{\alpha',\beta} \delta_{r,r'}$. The
tunneling rate $t$ is scaled out in the Hamiltonian.  Therefore, $m$ is
measured in units of $t$ and the fluctuation parameter $g$ is measured in units
of $t^2$.  This model, for $N=1$ and within a saddle-point approximation,
describes the jump in the conductivity characterising the QHT and displays
for the local DOS a characteristic semi-circle law \cite{zie1,zie3}. The
transition is driven either by changing the external field or, equivalently,
as will be argued in this work, by changing the chemical potential $\mu$.

In principle there can be three different types of randomness: a random
Dirac mass, a random energy and a random vector potential. We believe,
however, that the random mass alone correctly accounts for the effects of
the level-fluctuations in the single quantum dots. The random vector
potential does not break the time reversal symmetry of the massless
Dirac fermions. This is a strong restriction which implies that the
case of the massless model has very special properties \cite{lud}. Most
remarkable is the fact that the average DOS is singular and obeys a power
law: $\langle \rho(E) \rangle {\sim} E^{\alpha}$ with $-1<\alpha<1$. This is 
an unphysical behaviour because the DOS should be finite at the Hall 
transition. A finite DOS exists for a random Dirac mass \cite{zie2,zie3} and 
for a random energy term \cite{fradkin}. (Ludwig et al. \cite{lud} obtained 
for a random Dirac mass the result of a vanishing DOS at the Hall transition, 
using a perturbative renormalization group calculation. Consequently, they
argued that only a combination of a random mass {\it and} a random vector 
potential can lead to a non-zero DOS. However, a non-perturbative approach 
gives a non-zero DOS already for a random Dirac mass \cite{zie2}, proportional
to $\exp (-\pi/g)$.) The random Dirac mass creates spontaneously a 
contribution proportional to the random energy (see Ref. \cite{zie2} and 
Section III.A of this article). This implies that including the random energy 
will not lead to a qualitative change of the effect of the random Dirac mass.

Results from previous calculations with this model will be recalled below; here
it suffices to notice that the Green's function $G=(H+i\epsilon\sigma_0)^{-1}$
has special properties because of the Lorentz invariance of the Dirac
theory. In particular we will make use in the following of the relation
$(G_{11,rr}^{\alpha\alpha})^* =-G_{22,rr}^{\alpha\alpha}$. 
Here, and in the following, $\sigma_0$ is a convenient notation \cite{glja} for 
the 2$\times$2 unit matrix. 

As a last remark about the model, we now clarify the role of the ``fixed''
magnetic flux per plaquette.  The physical parameter of the model is not
the magnetic flux, but the filling factor $\nu=n \Phi_0/B$ ($n$ is here the
density of particles, $B$ the external magnetic field).  $n$ can be written as
the total number of particles divided by the total area of the system. Since
the particles are non-interacting fermions, there are at most $N$ particles,
if $N$ is the number of lattice sites. The local DOS is symmetric around
$\mu=0$. This implies that there are $N/2$ particles at $\mu=0$. On the
other hand, the lattice can be divided into plaquettes of four lattice sites
each. The flux through each plaquette is $\Phi_0/2=a^2 B$, if $a^2$ is the
area of the plaquette. Taking into account that there are $N$ plaquettes
(in the thermodynamic limit, when boundary effects can be ignored), we can
write for the filling factor

\begin{equation} \nu=(N/2) \Phi_0/(N a^2 B)=(N/2) \Phi_0/(N \Phi_0/2)= 1.
\end{equation}

\noindent Hence, the filling factor is 1, regardless of all other model
parameters.

\subsection{Second Moment of the Local DOS}

According to standard Green's function theory, the local DOS (LDOS) is
obtained from

\begin{equation} \rho^{\alpha\alpha}(r,M)=\frac{1}{\pi}
Im [G_{11,rr}^{\alpha\alpha} +G_{22,rr}^{\alpha\alpha}]=
\frac{i}{2\pi} \Big[G_{11,rr}^{\alpha\alpha}+G_{22,rr}^{\alpha\alpha}
-(G_{11,rr}^{\alpha\alpha})^*-(G_{22,rr}^{\alpha\alpha})^*\Big].
\end{equation}

\noindent For the tight-binding model without magnetic field this implies that
one has to evaluate the two-particle Green's function with opposite signs of
the frequency (advanced and retarded Green's functions) in order to get the
second moment of the LDOS \cite{efet,efpr}. This is the same problem as in
the evaluation of the conductivity. In the model with half a flux quantum per
plaquette, however, we can use the relation $(G_{11,rr}^{\alpha\alpha})^*=
-G_{22,rr}^{\alpha\alpha}$ to write

\begin{equation} \rho^{\alpha\alpha}(r,M)=\frac{i}{\pi}
(G_{11,rr}^{\alpha\alpha} +G_{22,rr}^{\alpha\alpha}), \end{equation}

\noindent and consequently

\begin{equation}
\rho^{\alpha\alpha}(r,M)\rho^{\beta\beta}(r,M)=-\frac{1}{\pi^2}
(G_{11,rr}^{\alpha\alpha}+G_{22,rr}^{\alpha\alpha})
(G_{11,rr}^{\beta\beta}+G_{22,rr}^{\beta\beta}).  \end{equation}

\noindent In this way we set out to evaluate

\begin{eqnarray} M_2^{\alpha\beta}&=&\langle\rho^{\alpha\alpha}(r,M)
\rho^{\beta\beta}(r,M)\rangle_{\delta M}
-\langle\rho^{\alpha\alpha}(r,M)\rangle_{\delta M}
\langle\rho^{\beta\beta}(r,M)\rangle_{\delta M} \nonumber
\\ &=&-\frac{1}{\pi^2} \left ( \langle G_{11,rr}^{\alpha\alpha}
G_{11,rr}^{\beta\beta} \rangle_{\delta M} - \langle G_{11,rr}^{\alpha\alpha}
\rangle_{\delta M} \langle G_{11,rr}^{\beta\beta} \rangle_{\delta M} + \langle
G_{11,rr}^{\alpha\alpha}G_{22,rr}^{\beta\beta}\rangle_{\delta M} - \langle
G_{11,rr}^{\alpha\alpha} \rangle_{\delta M} \langle G_{22,rr}^{\beta\beta}
\rangle_{\delta M} \right ) +[1\longleftrightarrow 2].  \end{eqnarray}

\section{ Functional Integral Representation }

By means of a standard representation, we write, quite generally \cite{zie3}

\begin{equation} G_{rr,jj}^{\alpha\alpha}
=-i\int\chi_{rj}^{\alpha}{\bar\chi}_{r,j}^\alpha\exp(-S_1)
\prod_r d\Phi_r d{\bar \Phi_r}
\equiv-i\langle\chi_{rj}^{\alpha}{\bar\chi}_{r,j}^\alpha\rangle_S,
\end{equation}

\noindent and consequently

\begin{equation} G_{rr,jj}^{\alpha\alpha}G_{rr,kk}^{\beta\beta}
=\int\chi_{r,j}^{\alpha}{\bar\chi}_{r,j}^\alpha\Psi_{r,k}^{\beta}
{\bar\Psi}_{r,k}^\beta\exp(-S_1)\prod_r d\Phi_r d{\bar \Phi_r}
\equiv\langle\chi_{r,j}^{\alpha}{\bar\chi}_{r,j}^\alpha\Psi_{r,k}^{\beta}
{\bar\Psi}_{r,k}^\beta\rangle_S, \label{corr} \end{equation}

\noindent with the supersymmetric action (sum convention for $\alpha$)

\begin{equation}
S_1=-i\sigma_\epsilon[(\Phi,(H_0+\epsilon\sigma_0){\bar\Phi})+\sum_r\delta
M_{r}^{\alpha\alpha'}
(\Phi_{r}^{\alpha'}\cdot\sigma_3{\bar\Phi}_{r}^{\alpha})], \end{equation}

\noindent where $\sigma_\epsilon ={\rm sign}(\epsilon)$ and the field
$\Phi_{r,j}^\alpha=(\Psi_{r,j}^\alpha,\chi_{r,j}^\alpha)$.  The first
component is Grassmann and the second complex.  We notice that the
normalization of the functional integral in (\ref{corr}) is due to the
combination of Grassmann and complex fields.  Averaging with Gaussian
distributed fluctuations in the ``masses'' $M^{\alpha\beta}_r$ yields
$\exp(-S_2)=\langle \exp(-S_1)\rangle_{\delta M}$, with

\begin{equation} S_2=-i\sigma_\epsilon(\Phi,(H_0+\epsilon\sigma_0){\bar\Phi})+
{g\over N}\sum_r (\Phi_{r}^\alpha\cdot\sigma_3 {\bar\Phi}_{r}^\alpha)^2.
\label{quartact} \end{equation}

\noindent Thus we have derived an effective field theory for $\Phi$ which
serves as a generating functional for the averaged Green's function.  It is
important to notice that {\it not only} $\delta M$ creates the fermion-fermion
interaction in (\ref{quartact}) but that also other types of randomness can
do this job. For instance, the interaction can also be created by a term
which couples to a matrix field ($\mu=1,...,4$ includes the complex and
Grassmann components):

\begin{equation} \exp\Big[-{g\over N}\sum_r (\Phi_{r}^\alpha\cdot\sigma_3
{\bar\Phi}_{r}^\alpha)^2\Big]=\int\exp\Big[-(N/g)\sum {\hat Q}_{r;\mu,\mu'}
(\sigma_3)_{\mu'}{\hat Q}_{r;\mu',\mu}(\sigma_3)_{\mu} -i\sum {\hat
Q}_{r;\mu,\mu'}\Phi_{r,\mu'}^{\alpha}{\bar\Phi}_{r,\mu}^{\alpha} \Big]{\cal
D}[{\hat Q}], \end{equation}

\noindent with the supermatrix

\begin{equation} {\hat Q}=\pmatrix{ Q&{\bar\Theta}\cr \Theta&-iP\cr }.
\end{equation}

\noindent The field $\Phi$ appears only in a quadratic form on the right
hand side.  Therefore, it can be integrated out. This leads to

\begin{equation} \langle G_{rr,jj}^{\alpha\alpha}\rangle_{\delta M}
=\int{\cal G}_{11,jj}\exp(-NS_3){\cal D}[Q,P,\Theta] \equiv\langle {\cal
G}_{11,jj}\rangle_Q, \end{equation}

\noindent and consequently

\begin{eqnarray} \langle
G_{rr,jj}^{\alpha\alpha}G_{rr,kk}^{\beta\beta}\rangle_{\delta M} & =
& \int(-{\cal G}_{12,kj}{\cal G}_{21,jk}\delta_{\alpha,\beta} +{\cal
G}_{11,jj}{\cal G}_{22,kk}) \exp(-NS_3){\cal D}[Q,P,\Theta] \nonumber\\ & &
\equiv\langle -{\cal G}_{12,kj}{\cal G}_{21,jk}\delta_{\alpha,\beta} +{\cal
G}_{11,jj}{\cal G}_{22,kk}\rangle_Q, \end{eqnarray}

\noindent where we have defined

\begin{equation} {\cal G}=\pmatrix{ G_0^{-1}-2\tau Q\tau
&-2\tau{\bar\Theta}\tau\cr -2\tau\Theta\tau & G_0^{-1}+2i\tau P\tau \cr
}^{-1}_{rr}.  \end{equation}

\noindent Here $G_0=(H_0+i\epsilon\sigma_0)^{-1}$, $\tau=\sqrt{\sigma_3}$,
$H_0\equiv\langle H\rangle=(\sigma_1\nabla_1+\sigma_2\nabla_2)+\sigma_3m$ and

\begin{equation} S_3= {1\over g} Trg\Big(\pmatrix{
Q&{\bar\Theta}\cr \Theta&-iP\cr }^2\Big) +\log\Big[ detg
\pmatrix{i(H_1+i\epsilon\sigma_0-2\tau Q\tau)&-2i\tau{\bar\Theta}\tau\cr
-2i\tau\Theta\tau&i(H_1+i\epsilon\sigma_0+2i\tau P\tau)\cr }\Big],
\end{equation}

\noindent where we have introduced the graded (or \cite{efet} super-) trace
and determinant. This implies for the local DOS

\begin{equation} {1\over
N}\sum_\alpha\langle\rho^{\alpha\alpha}(r,M)\rangle_{\delta M} ={1\over
{N\pi}}\sum_\alpha\sum_j\langle\chi_{rj}^\alpha{\bar\chi}_{rj}^\alpha
\rangle_S ={i\over {g\pi}}\sum_j\langle[\tau Q_r\tau]_{jj}\rangle_Q.
\label{ldos} \end{equation}

\noindent For the correlation of the local DOS, it follows that

\begin{equation} M_2^{\alpha\beta} =\frac{1}{\pi^2} \sum_{j,k=1}^2\Big[\langle
{\cal G}_{12,kj} {\cal G}_{21,jk}\delta_{\alpha, \beta}-{\cal G}_{11,jj}{\cal
G}_{22,kk}\rangle_Q+\langle {\cal G}_{11,jj} \rangle_Q\langle {\cal
G}_{11,kk}\rangle_Q\Big].  \end{equation}

\subsection{Saddle Point Approximation}

The number of levels $N$ appears in front of the action.  Thus the effect of
level fluctuations for $N\to\infty$ can be evaluated within a saddle point
(SP) approximation \cite{zie3}. The SP equation reads

\begin{equation} {\delta\over\delta Q}\big\lbrack {1\over g}Tr Q^2+
\log\det(G^{-1}_0-2\tau Q\tau)\big\rbrack=0.  \label{spe} \end{equation}

\noindent A second SP equation appears from the variation of $P$ by
replacing $Q\to -iP$. As an ansatz we take a uniform SP solution $Q_0 =-i P_0
=-(1/2)[i\eta\sigma_3+m_s\sigma_0]$.  Then (\ref{spe}) leads to the conditions
$\eta =\eta g I$, $m_s=-mgI/(1 +gI)$ with the integral $I=\int\lbrack
(m+m_s)^2 +\eta^2+k^2\rbrack^{-1}d^2k/2\pi^2$.  There is both a trivial and
a nontrivial solution, $\eta=0$ and $\eta\ne0$.  The latter is possible for
$m^2<m_c^2$ with $m_c=2\exp(-\pi/g)$. This means the level fluctuations shift
the energy $\epsilon\to\epsilon+\eta$ and the Dirac mass $m\to{\bar m}=m+m_s$,
where $\eta(m)$ and $m_s(m)$ are solutions of the SP equations (\ref{spe}).
The sign of $\eta$ is fixed by the condition that $\eta$ must be analytic
in $\epsilon$; this leads to ${\rm sign}(\eta)={\rm sign}(\epsilon)$. It is
important to notice that $\eta$ is proportional to the energy and {\it not}
to the mass. That means, the random Dirac mass creates spontaneously a 
contribution which would be expected from a random energy term. This fact 
indicates that an additional random energy term may not qualitatively change 
the properties of the random Dirac mass.

The average local DOS can now be directly calculated from (\ref{ldos}) in SPA

\begin{equation} \langle\rho^{\alpha\alpha}(r,M)\rangle_{\delta
M}\approx{\eta\over\pi g} ={1\over2\pi g}\sqrt{m_c^2-m^2}\Theta(m_c^2-m^2),
\label{dos} \end{equation}

\noindent where $\Theta(x)$ is the Heaviside step function. We see that a
semi-circle law is reproduced. In the following we will consider only the
regime where the average local DOS is nonzero, i.e., $m^2< m_c^2$.

\subsection{Gaussian Fluctuations}

In order to evaluate the second moment of the local DOS the Gaussian
fluctuations around the SP must be calculated.  Since $Q$, $P$ and $\Theta$
are $2\times2$ matrices, the fluctuations can also be parametrised as
4-component vector fields: $q_1=\delta Q_{11}$, $q_2=(\delta Q_{12}+\delta
Q_{21})/2$, $q_3=-i(\delta Q_{12}-\delta Q_{21})/2$, $q_4=\delta Q_{22}$
with analogous definitions for $p_1,...,p_4$ and the Grassmann field
$\psi_1,...,\psi_4$ with $\psi_2=(\Theta_{12}+\Theta_{21})/2$ and
$\psi_3=-i(\Theta_{12}-\Theta_{21})/2$.  The action of the Gaussian
fluctuations reads, in Fourier representation \cite{zie3}

\begin{equation} S\approx \int\sum_{\mu,\mu'=1}^4 ({\bf
I}_k)_{\mu,\mu'}(q_{k,\mu} q_{-k,\mu'}+p_{k,\mu}p_{-k,\mu'}+2{\bar
\psi}_{k,\mu}\psi_{-k,\mu'})d^2k, \end{equation}

\noindent with the fluctuation matrix ${\bf I}_k$. The stability matrix reads

\begin{equation} {\bf I}(k')=\pmatrix{ I_{11}&I_{12}&I_{13}&I_{14}\cr
-I_{12}^*/\rho&I_{22}&I_{23}&\rho I_{12}\cr -I_{13}^*/\rho&I_{23}&I_{33}&\rho
I_{13}\cr I_{14}^*&-I_{12}^*&-I_{13}^*&I_{11}^*\cr }, \end{equation}

\noindent with $\rho=\mu/\mu^*$ and $\mu=m/2+i\eta$.  In particular, for a
vanishing wave vector we have

\begin{equation} {\bf I}(k'=0)=\pmatrix{
1/g-{\mu^*}^2/2\pi|\mu|^2&0&0&1/g-1/2\pi\cr 0&2/g-1/\pi&0&0\cr
0&0&2/g-1/\pi&0\cr 1/g-1/2\pi&0&0&1/g-\mu^2/2\pi|\mu|^2\cr }.  \label{I0}
\end{equation}

\noindent Thus the Gaussian fluctuations are always massive except for the
critical points $m=\pm m_c$. This reflects the discrete symmetry of the
Hamiltonian (\ref{diracham}) (for a more detailed discussion of the symmetry
properties see Ref. \cite{zie3}). For $k'\sim0$ we obtain

\begin{eqnarray} I_{11}&=&1/g-{\mu^*}^2/2\pi|\mu|^2
\nonumber \\ I_{12}&=&{\mu^*\over2\pi|\mu|}{k'_2\over|\mu|}
\nonumber \\ I_{13}&=&{\mu^*\over2\pi|\mu|}{k'_1\over|\mu|}
\nonumber \\ I_{14}&=&1/g-1/2\pi+o({k'}^2) \nonumber \\
I_{23}&=&{1\over2\pi}{k'_1k'_2\over|\mu|^2} \nonumber \\ I_{22/33}&=&2/g-1/\pi,
\end{eqnarray}

\noindent that is, $I_{14}$ is real.

\subsection{Expansion of ${\cal G}$}

Retaining only terms up to the second order, we have

\begin{equation} \langle {\cal G}_{12,kj}{\cal G}_{21,jk}\delta_{\alpha,\beta}-
{\cal G}_{11,jj}{\cal G}_{22,kk}\rangle_Q+\langle {\cal
G}_{11,jj} \rangle_Q\langle {\cal G}_{11,kk}\rangle_Q
\approx4\delta_{\alpha,\beta}\langle[{\cal G}_0\tau\Theta\tau{\cal G}_0
]_{rr,kj}[{\cal G}_0\tau{\bar\Theta}\tau{\cal G}_0]_{rr,jk}\rangle_{\delta
Q}, \end{equation}

\noindent with

\begin{equation} {\cal G}_0=\pmatrix{ (H_0-2\tau
Q_0\tau+i\omega\sigma_0)^{-1}&0\cr 0&(H_0-2\tau
Q_0\tau+i\omega\sigma_0)^{-1}\cr }.  \label{matrix} \end{equation}

\noindent Now we have to perform the Grassmann integrations, ending up with

\begin{equation}
\langle\Theta_{j'j'',r'}{\bar\Theta}_{k'k'',r''}\rangle_{\delta
Q} =U_{j'j'',\nu}U_{k'k'',\nu'}^T
\langle\psi_{\nu,r'}{\bar\psi}_{\nu',r''}\rangle_{\delta Q}
=(1/2)U_{j'j'',\nu}U_{k'k'',\nu'}^T{\bf I}^{-1}_{r'r'',\nu\nu'}, \end{equation}

\noindent with the unitary matrix

\begin{equation} U={1\over\sqrt{2}}\pmatrix{ 1&0&0&0\cr 0&1&-i&0\cr 0&1&i&0\cr
0&0&0&1\cr }.  \end{equation}

\noindent The stability matrix (\ref{I0}) has one zero eigenvalue if $\mu$
is real, i.e.  for $m=\pm m_c$. Consequently, there is a singularity in the
Fourier components of ${\bf I}(k)$ at $k=0$. The leading behavior in the
$k\sim0$ asymptotics of ${\bf I}^{-1}$ can be extracted as

\begin{equation} \pmatrix{ \langle\Theta_{11}{\bar\Theta}_{11}\rangle_{\delta
Q}& \langle\Theta_{11}{\bar\Theta}_{22}\rangle_{\delta
Q}\cr \langle\Theta_{22}{\bar\Theta}_{11}\rangle_{\delta Q}&
\langle\Theta_{22}{\bar\Theta}_{22}\rangle_{\delta Q}\cr }={1\over2}\pmatrix{
{\bf I}^{-1}_{11}&{\bf I}^{-1}_{14}\cr {\bf I}^{-1}_{41}&{\bf
I}^{-1}_{44}\cr } ={-1/4\over1-\cos(2x)+(g/2\pi|\mu|^2)\cos(x) k^2} \pmatrix{
-2\pi+e^{2ix}g&2\pi-g\cr 2\pi-g&-2\pi+e^{-2ix}g\cr }, \end{equation}

\noindent where $\mu^*=|\mu|e^{ix}$. We consider weak fluctuations: higher
order contributions in $g$ in the numerator and in the coefficient of $k^2$
have been neglected. We make use of the following approximation

\begin{equation} (H_0-2\tau Q_0\tau+i\omega\sigma_0)^{-1}_{rr'}
\approx\delta_{r,r'}\int (|\mu|^2+k^2)^{-1}d^2k/(2\pi)^2\pmatrix{ \mu^*&0\cr
0&-\mu\cr }=(1/4\pi)\delta_{r,r'}\log(1+\lambda^2|\mu|^{-2})\pmatrix{
\mu^*&0\cr 0&-\mu\cr }, \end{equation}

\noindent where $\lambda$ is a UV cut-off. For simplicity, we assume in the
following $\lambda=1$.  Then the second moment of the local DOS reads

\begin{equation} M_2^{\alpha\beta}=\delta_{\alpha\beta}\pi^{-2}
\sum_{l,l'}\gamma_l^2\gamma_{l'}^2\tau_l^2\tau_{l'}^2 \int
K_{l,l'}(k)d^2k/(2\pi)^2, \end{equation}

\noindent with $K_{1,1}={\bf I}^{-1}_{11}$, $K_{1,2}={\bf I}^{-1}_{14}$,
 $K_{2,1}={\bf I}^{-1}_{41}$ and $K_{2,2}={\bf I}^{-1}_{44}$
and $\gamma_1=\mu^*\log(1+ |\mu|^{-2})$, $\gamma_2=-\gamma_1^*$.  Furthermore,

\begin{eqnarray} M_2^{\alpha\beta}
&=&\delta_{\alpha,\beta}\pi^{-2}\Big[2Re{\mu^*}^4(2\pi-e^{2ix}g)-2|\mu|^4
(-2\pi+g)\Big]C\int[1-\cos(2x) +(g\cos(x)/2\pi|\mu|^2)k^2]^{-1}d^2k/(2\pi)^2,
\end{eqnarray}

\noindent where $C=[(1/4\pi)\log(1+|\mu|^{-2})]^4/2$. The factor in front
of the integral reads also

\begin{equation} \pi^{-2}\Big\{4\pi[\cos(4x)+1]-2g[\cos(2x)+1]\Big\} |\mu|^4.
\end{equation}

\noindent Integration with respect to $k$ gives

\begin{equation} \int[1-\cos(2x)+(g\cos(x)/2\pi|\mu|^2)k^2]^{-1}
kdk/2\pi \sim -{|\mu|^2\over 2g\cos(x)}\log\Big\{ {1-\cos(2x)\over1
-\cos(2x)+(g\cos(x)/2\pi|\mu|^2) }\Big\}.  \end{equation}

\noindent Then

\begin{equation} M_2^{\alpha\beta}\approx -\delta_{\alpha,\beta}C{4|\mu|^6
[\cos(4x)+1]\over g\cos(x)}\log\Big\{ {1-\cos(2x)\over1-\cos(2x)+(g\cos(x)/2
\pi|\mu|^2) }\Big\}.  \end{equation}

\section{discussion}

Making use of the evaluation described above, we can now display and discuss
the dependence of $M_2$ on the effective chemical potential $m=\mu-t'$ and
on the amount of chaos or disorder in each single dot, $g$.

According to the average DOS (\ref{dos}) there is a narrow band of bulk
electronic states. Moreover, there are also edge states which carry the Hall
current for $m>m_c$. In the following we will discuss only the bulk states.
At the band center the second moment of the LDOS is $M_2^{\alpha\beta}=
\delta_{\alpha\beta}(m_c/2g)^4/\pi$. The ratio $\sqrt{M_2}/\rho$ at $m=0$
then is $\sqrt{\pi}m_c/2g$.

Fig. 1 shows the dependence of $M_2$ (suitably scaled) on the parameters $m$
and $g$. This figure, obtained for typical values of the parameters, displays
an interesting dramatic increase as a function of $m$ for the second moment
$M_2$ of the local DOS as the edge of the energy band, $m=m_c$, is approached.
This (logarithmic) divergence of $M_2$ at the edge of the band indicates a
power law singularity for the fluctuations near the QHT.  This is in agreement
with the results of the $2+\epsilon$ expansion \cite{wegner1,pruisken}. Also,
notice that, as they should, the fluctuations decrease as randomness decreases,
except in the vicinity of $m=m_c$. To our knowledge, these results, obtained
for a realistic lattice of ``artificial atoms'',  have not been discussed
before (but see, however, below). The data in Fig. 1 show quite clearly
the presence of important fluctuations in the local DOS as a consequence of
single-dot level fluctuations, the fluctuations becoming perhaps critical
at the edge of the band.

The level fluctuations for an isolated quantum dot are described by a single
random matrix $\sigma_3M^{\alpha\beta}$ which remains from the Hamiltonian
of Eq.(\ref{diracham}). The random matrix has, in contrast to the full
Hamiltonian (\ref{diracham}), a continuous unitary symmetry. This generates
in the $N\to\infty$ limit a saddle point manifold which we have to integrate
out.  The remaining functional is a zero-dimensional nonlinear sigma model
\cite{efpr} for the evaluation of the fluctuations of the DOS. The cross-over
from the fully coupled quantum dots to isolated dots can be achieved by
decreasing the tunneling rates $t$ and $t'$. Since $t$ was scaled out in
the Hamiltonian, we must study for this purpose the rescaled mass $m/t$
and the rescaled disorder parameter $g/t^2$. Sending $t\to0$, $m_c$ goes
like $2t/\sqrt{\exp(2\pi t^2/g)-1}\to \sqrt{2g/\pi}$. Therefore, the level
fluctuations create a broad ``band'' which is surpressed by the tunneling
term to a much narrower band. Moreover, the fluctuations of the local DOS
are also reduced by the tunneling between the quantum dots.

Notice that the strength of the fluctuations in the local DOS depends on the
parameter $m=\mu-t'$. This requires tuning the value of the chemical potential
$\mu$ or of the next-near-neighbour hopping $t'$. Both could be achieved in
experimentally realizable QDA by means of a varying gate voltage or a
suitable design in the gate structure and in its voltages.

It is important to discuss, at this point, the relevance of our calculation
for perspective experimental observations on QDA. To our knowledge the
local DOS can be accessed experimentally by two possibilities: the Knight
shift of the NMR (nuclear magnetic resonance) lineshape due to the electron
polarization \cite{slichter} and the SET (single electron tunneling) technique
\cite{schmidt}. From these measurements one can gain access to the statistics
of the local DOS mesoscopic fluctuations.

\subsection{ NMR Knight shift spectroscopy }

We begin by discussing the NMR Knight shift technique, which has recently
witnessed considerable revival both for mesoscopic semiconductor structures
and for high-temperature superconductors. For zero-dimensional mesoscopic
structures (single dots, small metallic particles or random aggregates of
these) the Knight shift method has been discussed by Efetov and Prigodin
\cite{efet,efpr} and by Fal'ko and Efetov \cite{faev}. For the one-dimensional
disordered metal a related discussion for the connection between the
local DOS and the Knight shift can be found in the work of Altshuler and
Prigodin \cite{alpr}. Here we give a formulation for the Knight shift in a
2D QDA. We adapt the treatment for the Knight shift in a traditional solid
(like, e.g. Li) \cite{slichter} to the situation in which a (2D) solid of
``artificial atoms'' is considered. Assume that a tight-binding description
of quasielectron tunneling (or hopping) is appropriate (as advocated in the
calculation of the previous Sections) and that a single-dot confining potential
approximately of the square-well type can be used to describe the single
``atomic'' levels $\{E_{\alpha}, \varphi_{\alpha}({\bf r})\}$ of a single
dot. Then a ``band-structure'' approach becomes a realistic calculation of
the energy levels of the entire 2D QDA made up of these ``atoms'' and we can
envisage the creation of ``bands'' of energy states from each ``atomic'' level
$E_{\alpha}$. We label the 2D QDA energy levels, therefore, $E_{\alpha}({\bf
k})=E_{\alpha}+\sum_{\delta}e^{i{\bf k}\cdot{\delta}}$ and these will normally
overlap due to the closeness of the levels $E_{\alpha}$ of the single-dots. By
building the many-electron wave-function in the normal way, the hyperfine
interaction Hamiltonian can be averaged over the electronic degrees of freedom
to give the contribution to the nuclear Hamiltonian arising from the electron
polarization in the QDA that ultimately leads to the Knight shift formula

\begin{equation} \frac{\Delta\omega({\bf r})}{\omega_0}=\frac{8\pi}{3}\int
d{\varepsilon} \chi (\varepsilon) \rho ({\bf r},\varepsilon) \langle
|u_{\varepsilon}({\bf r})|^2 \rangle |_{E_{\alpha}({\bf k})=\varepsilon},
\label{ks} \end{equation}

\noindent at the nuclear site ${\bf r}$. In the above formula, beside the
local DOS which is the subject of the present investigation, appear also
the single-dot wavefunction $u_{\varepsilon}({\bf r})$ (averaged over an
equal-energy surface) and the electron spin susceptibility

\begin{equation} \chi (\varepsilon)=\frac{1}{H_0}\gamma_e\hbar \frac{1}{2}
\left ( f(\varepsilon,-\frac{1}{2})-f(\varepsilon,\frac{1}{2}) \right
) {\approx} \frac{(\gamma_e\hbar)^2}{8k_BT}\frac{1}{\cosh^2 \left (
\frac{\varepsilon-E_F}{k_BT} \right ) }, \label{susc} \end{equation}

\noindent where the approximation holds at low temperatures (where the
Fermi distribution function $f(\varepsilon,m_s)$ can be expanded). Since the
low-temperature form of $\chi (\varepsilon)$ is delta-like, we essentially
get that the Knight shift is proportional to the local DOS, evaluated at the
Fermi level, however through a multiplicative constant that is in principle
also randomly distributed, in a way that is beyond the scope of the present
treatment. We conclude with the form

\begin{equation} \frac{\Delta\omega({\bf r})}{\omega_0}{\approx}\frac{8\pi}{3}
\langle |u_{\varepsilon}({\bf r})|^2 \rangle |_{\varepsilon=E_F} \chi (E_F)
\rho (E_F,{\bf r}), \end{equation}

\noindent for the expected Knight shift from a QDA. If we assume that the
wavefunction at the Fermi level has much smaller fluctuations than the local
DOS, then the moments of the local DOS can be extracted by looking at the
moments of the Knight shift itself.

\subsection{ Single-electron tunneling spectroscopy }

SET spectroscopy is well-known as a tool for the investigation of discrete
single- and few-electron energy states in quantum dots \cite{ashoori}.
Lerner and Raikh \cite{lera} have proposed to apply the technique to study the
weak-localization effects in disordered heterostructures, and the use of SET
spectroscopy for imaging the local DOS in a disordered (bulk) semiconductor
heterostructure has been recently reported \cite{schmidt}. Thus, SET
spectroscopy appears to be a promising tool for an experimental investigation
of local DOS fluctuations in a QDA.

Lerner and Raikh \cite{lera} show that, for the resonant-tunneling junction
geometry, the measurement of the conductance $G_t$ gives direct imaging of
the local DOS, $\rho$:

\begin{equation} G_t=\frac{\pi e^2 S \Gamma_0}{2k\hbar} \rho, \end{equation}

\noindent with $\Gamma_0$, $k$ and $S$ geometrical parameters. A more
complicated treatment ensues for a disordered material in the junction's
barrier. The conductance presents mesoscopic fluctuations at low temperatures
and one can show \cite{lera} that the second moments of local DOS and
conductance are related by the formula

\begin{equation} \frac{M_2(G_t)}{\langle G_t \rangle}{\propto}
\frac{M_2(\rho)}{\langle \rho \rangle}.  \end{equation}

\noindent However, Schmidt et al. \cite{schmidt} propose a double-barrier
tunneling experimental geometry which leads through the measurement of the
$I-V$ characteristic to the relation

\begin{equation} \delta G({\bf r}) =\delta( dI/dV ) \propto \delta ( d\rho({\bf
r},E)/dE ), \end{equation}

\noindent for the {\it local} fluctuations of the conductance $G$ of the
disordered sample. Thus, SET spectroscopy can provide information, in
principle, to local DOS statistical properties perhaps also in the case of
a QDA setup.

We mention, before concluding this subsection, the possibility of imaging the
local DOS for surface electronic states also by means of scanning-tunneling
microscopy (STM) \cite{stm}. This technique also relies on the measurement
of the differential conductance, $dI/dV$, at local sample's sites; from
this, the shape of the local DOS and its fluctuations' statistics can be
reconstructed. This technique is perhaps more suitable for 2D systems like
our QDA.

Whether from the Knight shift in NMR or from the conductance in SET
experiments, the measurement of the value of the local DOS in a real QDA
should allow (via, e.g. repeated measurements) for a statistics of its
fluctuations and the experimental determination of $M_2$. This will be an
indirect observation of the existence of level-fluctuations in single
quantum dots.

\subsection{ Beyond 2D: $d=2+\epsilon$, 1 and 0 dimensions }

In dimensions higher than $d=2$, there already exist calculations of
the second moment of the local DOS for the situation in which randomness
creates a mobility edge in the Anderson transition for a disordered metal
\cite{wegner}. These calculations are based on $2+\epsilon$ type expansions
on the $Q$-matrix non-linear $\sigma$ model and lead to the result that $M_2
\sim (E-E_C)^{-\mu_2}$ with $\mu_2=2+O(\epsilon)$. This shows the existence
of the possibility of a divergence for the second moment of the local DOS
near the mobility edge. Although the Anderson transition situation is not
exactly the same QDA problem treated in the present paper, nevertheless
these calculations show that a divergence of $M_2$ is to be expected.

In lower dimensions, we recall the work of Altshuler and Prigodin, for $d=1$
\cite{alpr}, and of Efetov and Prigodin, for $d=0$ \cite{efpr} (a single
dot in our picture), for the case of a disordered metal. In $d=1$, exact
diagrammatic techniques can be employed to work out the entire probability
distribution for the local DOS of a weakly disordered metal. The resulting
distribution depends strongly on the type (open or closed) of boundary
conditions. Finally, in $d=0$ the non-linear-$\sigma$-model calculations
involving supersymmetry of Ref. \cite{efet,efpr} can be employed.

\subsection{ Conclusions }

We have shown how, by means of the technique of Dirac fermions in $d=2$,
one can characterize the distribution of the fluctuations of the local DOS
in the proximity of an IQHE transition. We have restricted the calculation
to  characterize the second moment $M_2(\rho)$, but in principle also the
higher moments can be evaluated (with increasing technical difficulties).

We have made extensive use of the fact that in the Dirac fermions approach
one is dealing with a unique saddle point solution, unlike in the case of the
non-linear-$\sigma$-model approach. This is a consequence of the discrete
symmetry of the model of Ludwig et al. \cite{lud} for the case of a random
mass, which we use for our calculations. This simple structure of the saddle
point solution made the 2D calculation possible.

\begin{center} ACKNOWLEDGEMENTS \end{center}

This work was supported in part (G.J.) by EC contract No. ERB4001GT957255.

\begin{center} FIGURE CAPTIONS \end{center}

Figure 1. Second moment $M_2$ of the local density of states (LDOS) as function
of effective chemical potential, $m$ and strength of level fluctuations, $g$.

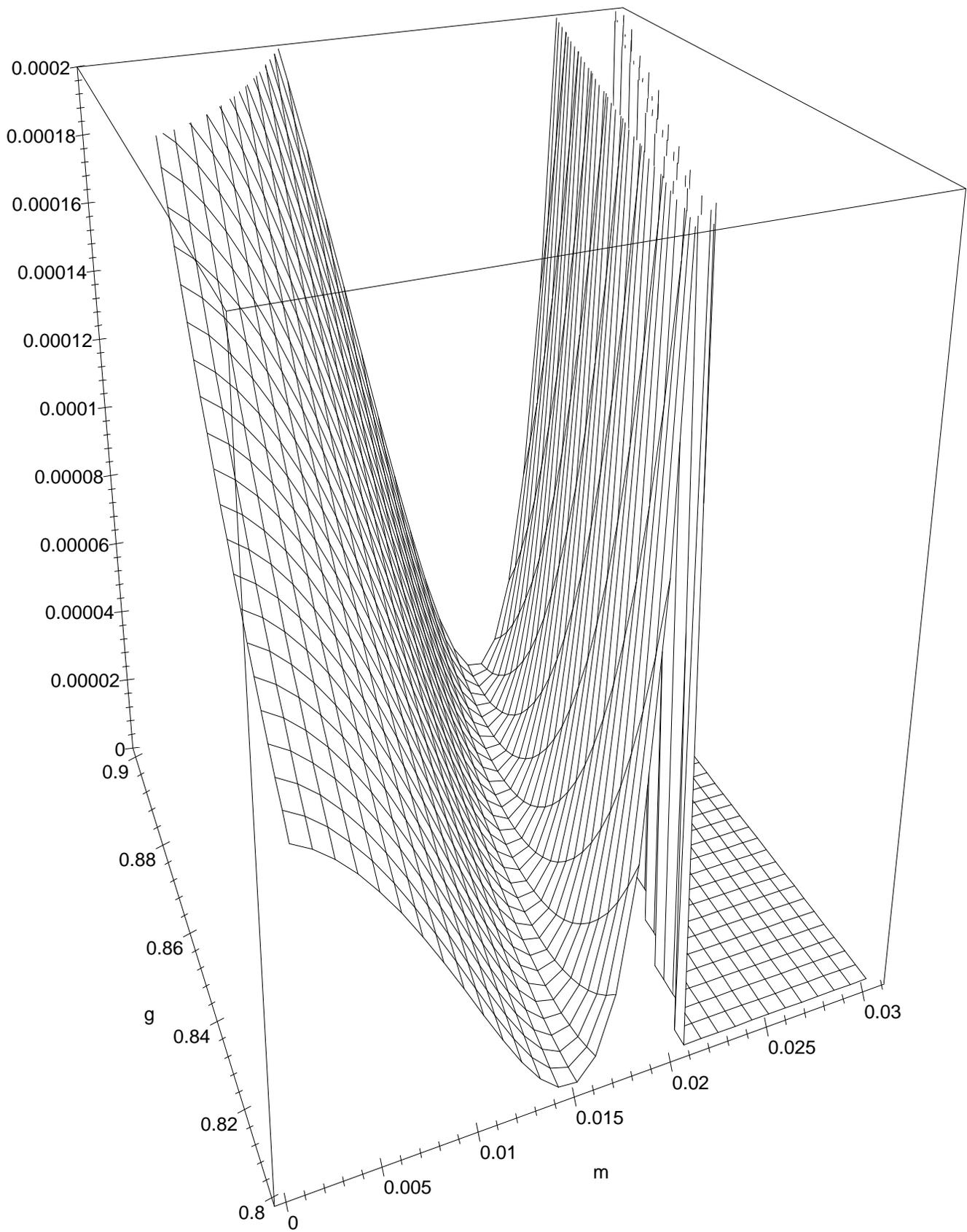
\end{document}